\documentclass{egpubl}
\JournalSubmission
\electronicVersion
\ifpdf \usepackage[pdftex]{graphicx} \pdfcompresslevel=9
\else \usepackage[dvips]{graphicx} \fi
\PrintedOrElectronic
\usepackage{t1enc,dfadobe}
\usepackage{egweblnk}
\usepackage{cite}

\usepackage{amsmath}
\usepackage{amsfonts}
\usepackage{amssymb}
\usepackage{caption}
\usepackage{subcaption}
\usepackage{xcolor}

\newcommand{\teaserfigwidth}{0.145\linewidth}

\newcommand{\colorbarwidth}{0.02\linewidth}

\pdfoutput=1

\title{Quantitative Analysis of Saliency Models}

\author[F.P. Tasse \& J. Kosinka \& N.A. Dodgson]
       {F.\,P. Tasse$^{1}$
        and J. Kosinka$^{1,2}$
        and N.\,A. Dodgson$^{1,3}$
        \\
         $^1$Rainbow Group, University of Cambridge, United Kingdom \\
         $^2$Scientific Visualization and Computer Graphics, University of Groningen, Netherlands \\
         $^3$Computer Graphics, Victoria University of Wellington, New Zealand
       }
       
\begin{document}

\teaser{
\centering
\begin{tabular}{cccccc}
Ground-truth \cite{Chen2012} & LS \cite{Shtrom2013} & MS \cite{Song2014} & CS \cite{Tasse2015} & PS (PCA-based) \\
\includegraphics[width=\teaserfigwidth]{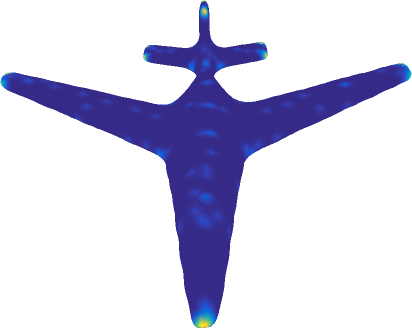} &
\includegraphics[width=\teaserfigwidth]{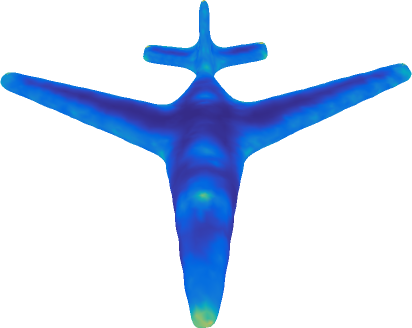} &
\includegraphics[width=\teaserfigwidth]{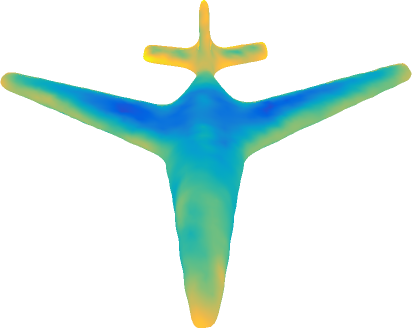} &
\includegraphics[width=\teaserfigwidth]{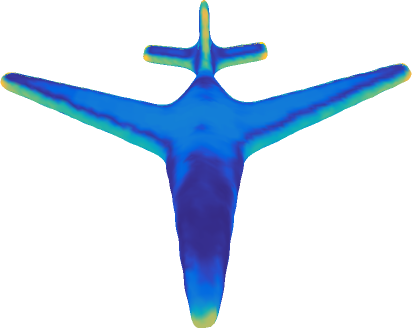} &
\includegraphics[width=\teaserfigwidth]{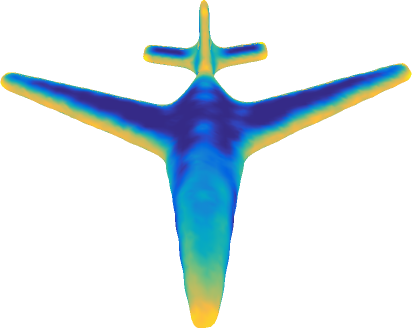} & \\
(AUC, NSS, LCC): & (\textbf{0.72}, 1.70, 0.76) & (0.66, 1.03, 0.63) & (0.69, 1.63, \textbf{0.78}) & (0.70, \textbf{1.71}, 0.75)  & \\

\includegraphics[width=\teaserfigwidth]{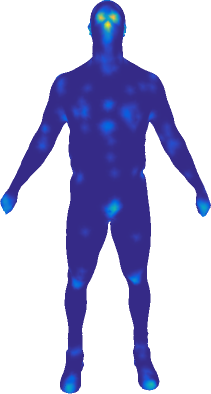}  &
\includegraphics[width=\teaserfigwidth]{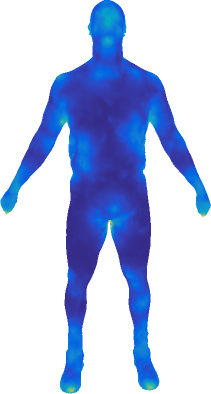}  &
\includegraphics[width=\teaserfigwidth]{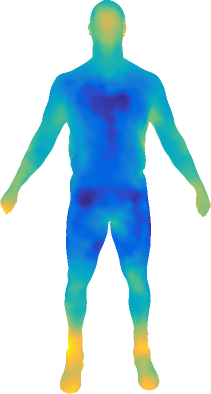} &
\includegraphics[width=\teaserfigwidth]{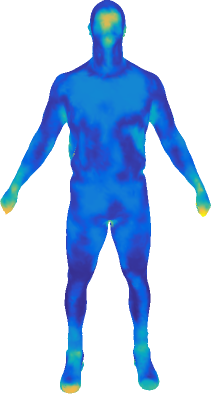} &
\includegraphics[width=\teaserfigwidth]{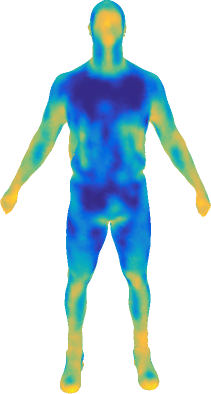} &
\includegraphics[width=\colorbarwidth]{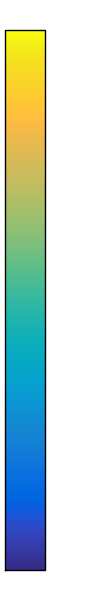} \\
(AUC, NSS, LCC): & (\textbf{0.62}, 0.70, 0.25) & (0.59, 0.53, 0.18) & (0.62, 0.88, 0.36) & (0.61, \textbf{1.05}, \textbf{0.38}) & \\
\end{tabular}

\caption{Examples of saliency maps generated by selected saliency models. The scores underneath each map are evaluation metrics that indicate how well the map compares to ground-truth. The metric for the top-performing saliency map is indicated in \textbf{bold}.}
 \label{fig:teaser}
 }
\maketitle

\begin{abstract}

Previous saliency detection research required the reader to evaluate performance qualitatively, based on renderings of saliency maps on a few shapes. This qualitative approach meant it was unclear which saliency models were better, or how well they compared to human perception. This paper provides a quantitative evaluation framework that addresses this issue. In the first quantitative analysis of 3D computational saliency models, we evaluate four computational saliency models and two baseline models against ground-truth saliency collected in previous work.
\end{abstract}

\begin{classification} % according to http://www.acm.org/class/1998/
\CCScat{Computer Graphics}{I.3.5}{Computational Geometry and Object Modelling}{Curve, surface, solid, and object representations}
\end{classification}

\section{Introduction}
We introduce three metrics for quantitative analysis and comparison of 3D saliency methods.

Despite recent interest in saliency detections of 3D surfaces, there are no evaluation metrics available, and no previous quantitative evaluation of 3D saliency models, making it impossible to objectively compare one method against another. Previous work evaluates results qualitatively, by showing 3D renderings of saliency maps of a few shapes to the reader. This approach does not objectively determine whether one saliency model is better than another. With the absence of a saliency evaluation benchmark, new techniques cannot be compared to previous methods on a common dataset. To the best of our knowledge, no previous work performs a quantitative evaluation of computational models to ground truth. We propose such a  quantitative evaluation, based on surface feature points acquired from users by Chen et al.~\cite{Chen2012}. Inspired by Judd et al.'s benchmark of computational models to predict human eye fixations in images~\cite{Judd2012}, we evaluate $4$ saliency models, and $2$ baseline models against ground truth 3D saliency on watertight meshes and simulated range scans.

Quantitative studies on shape saliency are few. Howlett et al. \cite{Howlett2005} empirically show, using eye fixations data collected from users, that mesh saliency exists and is useful in mesh simplification. Kim et al. \cite{Kim2010} compare mesh saliency \cite{Lee2005} to human eye movements on $5$ meshes, and show that a computational saliency models human eye fixations significantly better than chance. A larger study by Chen et al. \cite{Chen2012}, on $400$ models from Giorgi et al. dataset \cite{Giorgi2007},
collects ground-truth saliency data via salient point selection  by users. The study shows strong correlations between saliency as perceived by humans and mesh properties such as curvature. 

An unexplored area of empirical research on 3D saliency detection is a saliency benchmark that compares saliency models against one another and human performance. Benchmarks for 3D keypoint detection have been proposed \cite{Dutagaci2012}, as well as quantitative analysis that explain how mesh properties such as curvature are correlated with ground-truth saliency using information theory statistics \cite{Chen2012}. However, none of these works compare saliency performance using raw saliency maps. In contrast, there are well-established 2D image saliency benchmarks \cite{Judd2012, Borji2013}, which provide a framework for quantitative evaluation of different 2D saliency algorithms. Thus, we use available ground-truth saliency data \cite{Chen2012} and present evaluation metrics, inspired by the 2D case, to compare computational saliency models. 
%}

Our contributions are  three metrics for estimating the performance of saliency models, evaluated on six saliency models' ability to predict ground truth saliency, on a dataset of $400$ triangular meshes from the SHREC'07 Watertight models track~\cite{Giorgi2007} (SHREC07) and a synthetic dataset of $4800$ single-view 3D scans.

\section{3D Saliency methods}\label{sec:saliency_models}
We review recent computational saliency models, with a particular emphasis on methods evaluated in this paper.
\subsection{Saliency of large point sets ($\mathbf{LS}$) \cite{Shtrom2013}}
%We use our implementation of their technique to generate the saliency maps used in the evaluation. 
Shtrom et al. \cite{Shtrom2013} propose the first method that supports saliency detection on large points sets. Inspired by Leifman et al.'s \cite{Leifman2012} work on triangular meshes, saliency is a combination of point distinctiveness at two scales with point association, a function that assigns higher saliency to regions near foci of attention. Formally, saliency of a point $p_i$ is:
$$S(p_i) = \frac{1}{2}(D_{low}(p_i)+D_{high}(p_i))+\frac{1}{2}A_{low}(p_i),$$
where $D_{low}$ and $D_{high}$ are the low-level and high-level distinctiveness, and $A_{low}(p_i)$ is the point association. The point distinctiveness of $p_i$  is computed as the average dissimilarity between $p_i$ and other points. This dissimilarity measure between two points is the $\chi^2$ distance between their local descriptors, weighted by their Euclidean distance to give more influence to nearby points. Shtrom et al.\ use Fast Point Feature Histograms (FPFH) as local descriptors \cite{Rusu2009} using two scales, denoted here with $r_{low}$ and $r_{high}$, to compute local descriptors for low-level and high-level distinctiveness, respectively.  Finally, foci of attention used in $A_{low}$ are computed by taking $20\%$ of the points with highest low-level distinctness. $A_{low}$ attributes higher saliency to points near these foci of attention. This saliency model is described as generating plausible saliency maps for small and large point sets such as city scans.
 
\subsection{Mesh saliency via spectral processing ($\mathbf{MS}$) \cite{Song2014}}
%Source code provided by the authors.
Song et al. \cite{Song2014} propose a spectral-based approach, described as more robust compared to previous mesh saliency methods that focused on analysing changes in local vertex properties \cite{Pauly2003, Lee2005, Castellani2008, Unnikrishnan2008}.  The new approach uses spectral properties of a mesh at multiple scales using  the $n$ lowest frequencies of its log-Laplacian spectrum $L$. The log-Laplacian spectrum amplifies variation in the low-frequency parts of the Laplacian spectrum and detects the most `fundamental' saliencies. Single-scale saliency is computed by taking the absolute difference between $L$ and a locally averaged log-Laplacian spectrum $A$, and mapping it back to the spatial domain. This single-scale method captures globally salient regions but ignores local details. The authors address this issue by computing saliency on a group of smoothed meshes at scales $\{\epsilon^2, 2\epsilon^2, 3\epsilon^2, 4\epsilon^2, 5\epsilon^2\}$ for a given $\epsilon$.
Finally, multi-scale saliency is obtained by applying the logarithm operator to a smoothed summation of the absolute differences of saliency maps at consecutive scales. This method is described as being able to capture both globally important regions and local saliencies.

\subsection{Cluster-based point set saliency ($\mathbf{CS}$) \cite{Tasse2015}}
%Cluster-based point set saliency ($\mathbf{CS}$) \cite{Tasse2015}: Source code provided by the authors.
Tasse et al. \cite{Tasse2015} propose a cluster-based saliency model presented as being able to detect fine-scale saliency with better  time complexity. They segment point sets into $K$ clusters, and compute cluster saliency as a sum of cluster distinctiveness and spatial distribution. The point-level saliency is obtained by smoothing cluster-level saliency. Cluster distinctiveness is based on the mean FPFH of points belonging to that cluster, using a method similar to Shtrom et al.'s \cite{Shtrom2013} low-level distinctiveness. It compares a cluster descriptor to every other cluster, with more influence accorded to nearby clusters, so that a region is more distinctive if it is different from its surroundings. Cluster spatial distribution computes the spatial variance of geometrically similar clusters. Both heuristics capture both local and global saliencies. Local descriptors are computed at a single scale $r$, as opposed to Shtrom et al.'s two-scale method.

\subsection{Other methods}
Early saliency models compute a multi-scale representation of a mesh and observe how a local vertex property such as curvature, surface variation or normal displacement changes at different scales \cite{Pauly2003, Lee2005, Castellani2008, Unnikrishnan2008}. These methods are tightly linked to local properties which are not robust against noise and topological changes. 

Other saliency models achieve robustness and speed by first segmenting a mesh into patches represented by descriptors, followed by a ranking process that specifies patch distinctiveness \cite{Gal2006, Wu2013, Tao2014}. Patch distinctiveness may then be assigned to individual vertices by smoothing with  a Gaussian kernel. This distinctiveness computation can also be done with robust vertex descriptors but this comes with a higher computational cost \cite{Gelfand2005}. Leifman et al. \cite{Leifman2012} compute vertex distinctiveness at a fine and coarser scale, and combine this with the idea that vertices close to foci of attention, such as extremities, are salient. Saliency models discussed so far, with a few exceptions \cite{Pauly2003, Unnikrishnan2008}, require topological information and thus cannot support other shape representations such as point clouds. Recent saliency models described in the above section, such as Shtrom et al. and Tasse et al., focus on point sets. They fill a gap in the literature that has become important owing to the proliferation of low-range scanning devices that produce point sets that may not always be suitable for full reconstruction. Thus, our paper evaluates selected saliency models on both typical watertight meshes and simulated range scans. 

Few methods use learning for shape saliency detection. As part of their large-scale user study on mesh saliency, Chen et al. \cite{Chen2012} propose two regression models, based on the collected data. The first model is trained over the whole dataset, and the second  model only on meshes from the same class. The second model produces better results than the first. Supervised learning for saliency detection is challenging to apply to a wider range of datasets due to lack of training data. Shilane and Funkhouser \cite{Shilane2007} learn distinctive regions without a need for saliency training data, given a collection of meshes partitioned into classes. Meshes are split into regions and the saliency of a region corresponds to how well that region defines the mesh class and differentiates it from other classes. This method identifies large salient regions, such as the head of a four-legged animal, but does not detect finer-scale features. 

\section{PCA-based saliency}\label{sec:pca_based_saliency}
Both Tasse et al. \cite{Tasse2015} and Shtrom et al. \cite{Shtrom2013}

use FPFH to compute saliency, under the premise that this descriptor captures the geometry of a local neighbourhood and can be used to best compute similarity between neighbourhoods. It is robust to noise and sampling density. We propose a new, simple, saliency method where saliency is the absolute value of the descriptor projected onto the largest principal axis after mean centering.  It is interesting to investigate how this simple model compares with recent computational models described above. Before providing details on the analysis of PCA-based saliency and other methods, we first describe FPFH descriptors.% and their PCA analysis.

The FPFH of  a point $p$ is computed by taking a distance-weighted average of the Simplified Point Feature Histogram (SPFH) of nearby points within a support sphere of radius $r$. SPFH captures mean curvature around a point using a multi-dimensional histogram that captures the sampled surface variations. The SPFH of a point $p$ is a histogram of angular variations between $p$ and other points in its neighbourhood. Given a pair of points $p_i$ and $p_j$ with their respective unit normals $n_i$ and $n_j$, the angular variations between the two are computed by first  defining an orthonormal Darboux coordinate frame $(u,v,w)$ by
\begin{equation*}
u = n_i, \quad v = u \times \frac{p_j-p_i}{\|p_j-p_i\|}, \quad w  = u \times v.
\end{equation*}
The relative difference between $p_i$ and $p_j$ with their estimated normals 

is then encoded by the triple $(\alpha, \phi, \theta)$, where
\begin{equation*}
\alpha = v{\cdot}n_j, \quad \phi = u{\cdot}\frac{p_j-p_i}{\|p_j-p_i\|}, \quad \theta=\mathrm{arctan2}(w{\cdot}n_j, u{\cdot}n_j).
\end{equation*}

Note that we use the PCL library \cite{Rusu2011} to compute $(\alpha, \phi, \theta)$ and take their absolute values to get invariance to reflection \cite{Shtrom2013}. 

An 11-bin histogram is computed for each angular parameter, $\alpha$, $\phi$, $\theta$, and the three histograms are then concatenated to form a 33-dimensional vector.This produces a scale-invariant descriptor robust to noise, which is a requirement for handling point sets from range scans. If the input are unoriented point sets, the normal at a point is approximated by analysing the eigenvectors of the covariance matrix created from its nearest neighbours. To consistently orient normals, we propagate a seed orientation through a minimum spanning tree of the Riemann graph constructed over the point set, as proposed by Hoppe et al. \cite{Hoppe1992}. Given the above definition of SPFH, denoted by $\hat{H}(p)$, the final FPFH descriptor $H(p)$ is a 33-dimensional weighted average of SPFH of neighbouring points:
\begin{equation*}
H(p) = \hat{H}(p) + \frac{1}{|\mathcal{N}(p)|}\sum_{q \in \mathcal{N}(p)}{\frac{1}{\|p-q\|_2}}\hat{H}(q),
\end{equation*}
where $\mathcal{N}(p)$ is the set of neighbours $q$ of $p$ with $\|p-q\|<r$.

Our PCA-based saliency approach (\textbf{PS}) is a variation of the above FPFH-based models that dramatically reduces the amount of information that needs to be compared when processing a shape.
For each 3D model, we apply PCA to a large matrix consisting of the FPFHs of each point, and  retain only the strongest component. This component should capture the distinctiveness of each point. This can be seen as saliency detection purely based on geometric similarity, with any spatial consideration ignored. 

\section{Methodology}\label{sec:benchmark}

We present our setup for a quantitative analysis of computational saliency models. 

\subsection{Datasets} 
We use two datasets to evaluate saliency performance on typical watertight meshes and and simulated range scans.

\subsubsection{Watertight meshes}
Ground-truth saliency is obtained from Chen et al.'s user study \cite{Chen2012}, which comprises salient points, referred to by the authors as Schelling points. Users were asked to select points that were likely to be selected by other users. Each mesh in the SHREC07 dataset was annotated by at least $22$ participants. Chen et al. also compute a scalar field over a mesh by smoothing, with a Gaussian filter, the frequency with which each vertex was selected by all participants. We use this scalar field as ground-truth saliency.

\subsubsection{Simulated range scans}
In addition to mesh data, we are also interested in how saliency models deal with challenging data such as single-view point scans. Using the SHREC07 dataset above, we generated a large dataset of $4800$ synthetic scans by rendering $12$  range images from each of $400$ meshes, and converting the range images to point sets. Similarly to Sipiran et al.'s \cite{Sipiran2013} generation of range images, each mesh is enclosed in a regular icosahedron, and each vertex of the icosahedron is used as a camera position. Scanned points are intersections between the mesh and rays shot from the camera. These points are in the same object coordinate system as their base mesh, making it possible to map a point on a scan to a point on the base mesh and record their ground-truth saliency value. 

Because some of our tested saliency models require meshes as input \cite{Song2014}, we reconstruct partial meshes from points sets using the Greedy Projection Triangulation method available in the PCL library. Triangulation is performed by projecting the local neighborhoods of points along their normals and linking unconnected points. 

The above approach generates $4800$ single-view range scans with ground-truth data that can be used in our evaluation.

\subsection{Evaluation metrics}
% Inspired by MIT work "A Benchmark of Computational Models of Saliency to Predict Human Fixations" (http://dspace.mit.edu/handle/1721.1/68590)
% Methods for comparing scanpaths and saliency maps: strengths and weaknesses by Olivier Le Meur & Thierry Baccino
Saliency evaluation benchmarks in 2D images are well-established. We adapt evaluation scores used in these benchmarks to 3D saliency. Each saliency model is compared against ground-truth ($\mathbf{GS}$) using the following $3$ metrics: 

\paragraph*{Area under the ROC curve (AUC):} The Receiver Operating Characteristic (ROC) curve is obtained by thresholding the saliency map into a binary mask that separates positive samples (salient points) from negative samples (non-salient points) and, for different threshold values, plotting the true positive rate against the false positive rate. This metric is commonly used to compare saliency models in the $2D$ case \cite{Judd2012, LeMeur2013}. The ideal saliency model has AUC of $1.0$. AUC disregards regions with no saliency, and focuses on the ordering of the saliency values. Other, more selective, metrics are needed to support evaluation.

\paragraph*{Normalized scanpath saliency (NSS):} Also widely used in comparing 2D saliency maps to human eye fixations \cite{LeMeur2013}, NSS measures saliency values at fixation points along each user's eye scanpath. In our 3D case, we consider points selected by users as fixation points. For each participant, a NSS score is computed by a weighted sum of the computational saliency at points selected by the participant. The final NSS metric is the average over all participants. The higher this metric, the closer the evaluated computational model is to ground-truth since interest points selected by users should have large saliency values.

\paragraph*{Linear correlation coefficient (LCC):} This coefficient measures the strength of the linear relationship between two variables \cite{Borji2013}. The coefficient ranges between $-1$ and $1$, with values closer to $0$ implying a weak relationship. We use its absolute value as the metric score. With $X$ the ground-truth and $Y$ the saliency map under consideration, the correlation coefficient is 
\begin{equation*}
LCC(X,Y)=\frac{|\mathrm{cov}(X,Y)|}{\sigma_X\sigma_Y},
\end{equation*}
where $\sigma_X$ and $\sigma_Y$ are the standard deviation of $X$ and $Y$, respectively, and $\mathrm{cov}(X,Y)$ is the covariance between the two variables. When $\sigma_Y=0$, we set $LCC(X,Y)=0.$
Note that one of the things captured by LCC is whether two distributions have peaks at the same place, but this is limited by the fact that LCC is greatly influenced by the shapes of these peaks. Despite its limitations, we include LCC since it is a popular metric for measuring linear relationships between distributions, and other saliency metrics can help mitigate these limitations.

%\paragraph*{Earth mover distance ($EMD$)}: This measures the similarity between two probabilistic distributions, and is used by \cite{Judd2012} to compare 2D saliency models to fixation maps. If two distributions can each be viewed as a unit amount of dirt over a region, then $EMD$ is the minimum cost of turning one pile into the other, which is assumed to be the amount of dirt transported times the distance travelled. We use a recent efficient formulation of $EMD$ on discrete surface \cite{Solomon2014}.  The $EMD$ metric varies between $0.0$ and $1.0$. Smaller $EMD$ distance between a computational saliency distribution and ground-truth means that the two distributions have a large overlap.

\medskip
Computational models that are the closest to ground-truth have high AUC and NSS scores and high absolute value of LCC . We use the Wilcoxon rank-sum test \cite{Wilcoxon1945}, a non-parametric alternative to the two-samples t-test,  at a $0.05$ significance level  to report statistically significant differences between saliency performances of competing methods.

\begin{figure}[t]
\centering
\includegraphics[width=\linewidth]{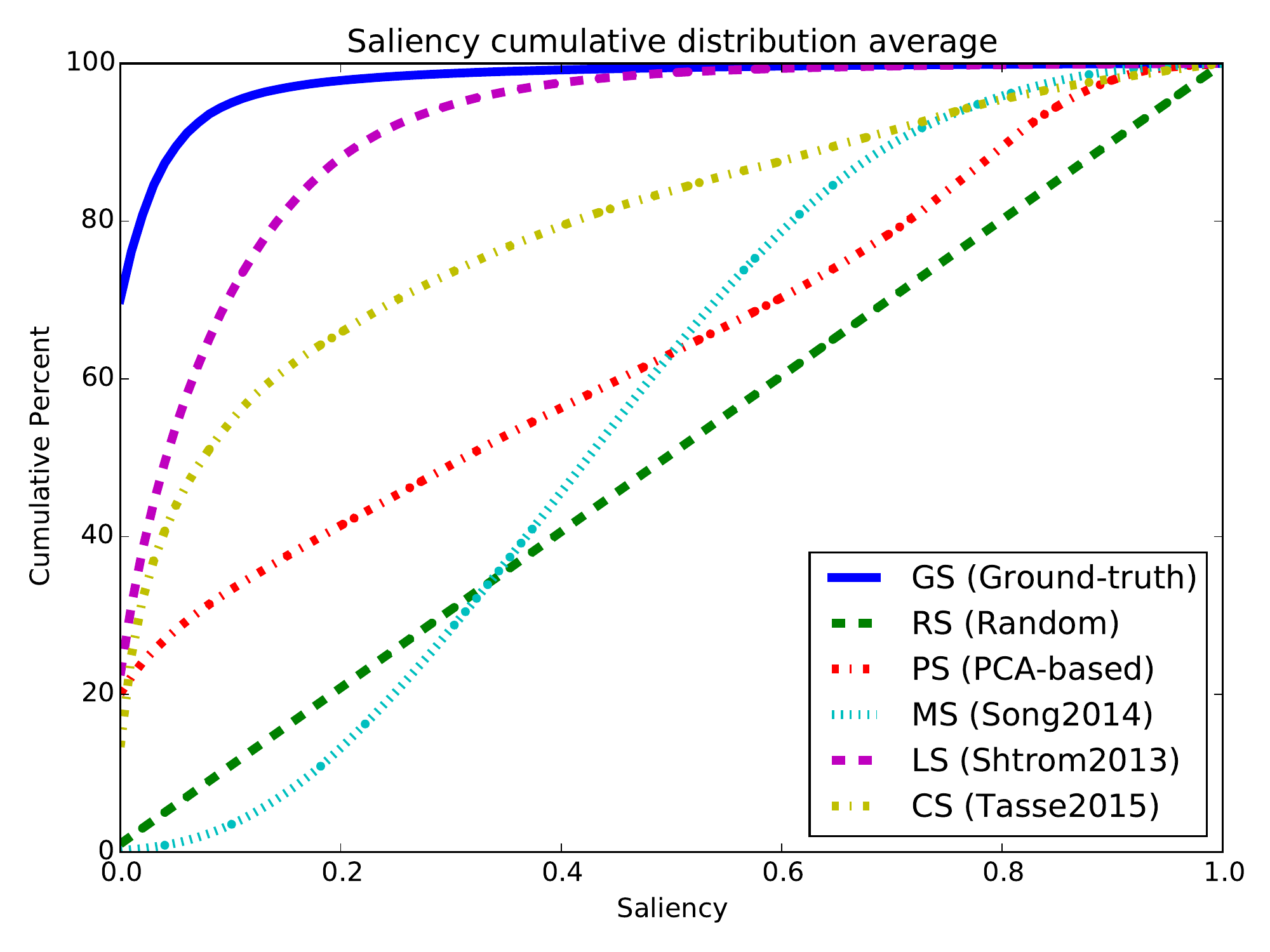}

\caption{Cumulative distribution functions (CDF) of maps generated by the evaluated saliency models. Each curve is the average of CDFs of SHREC07 saliency maps generated by the corresponding method. Random saliency has a CDF similar to that of uniform CDF, whereas maps generated by spectral mesh saliency have CDFs similar to the normal CDF. The graph shows that different saliency models produce different saliency distributions, hence the need to  adjust each shape saliency CDF to the average ground-truth CDF (histogram matching) before quantitative evaluation so that fair comparisons are made.}\label{fig:saliency_maps_avg_distribution}
\end{figure}

\begin{figure*}[ht]
\centering
\hspace*{\fill}
\includegraphics[width=0.15\linewidth]{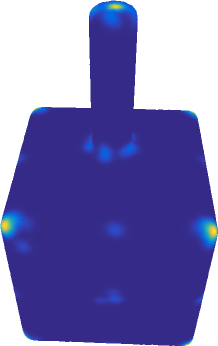} 
\hfill
\includegraphics[width=0.15\linewidth]{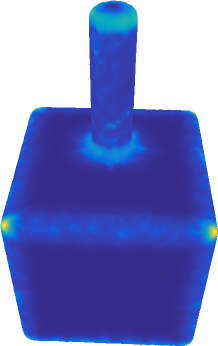}
\includegraphics[width=0.15\linewidth]{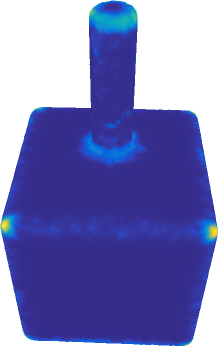}
\hfill
\includegraphics[width=0.15\linewidth]{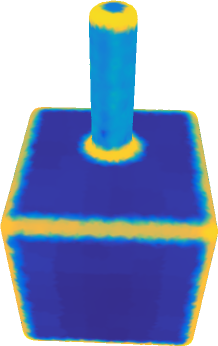}
\includegraphics[width=0.15\linewidth]{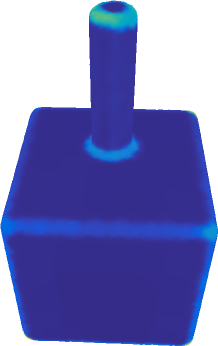}
 \hspace*{\fill}
\includegraphics[width=\colorbarwidth]{eval_results/parula_colorbar.png}

\caption{Saliency maps after their histograms were matched to ground truth. From left to right: Ground-truth, LS before matching, LS after matching, CS before matching, CS after matching. This matching allows a fair comparison of various saliency models. }\label{fig:maps_after_matching_hist}
\end{figure*}

\subsection{Selected saliency methods} 
We selected four computational models from the literature discussed in Section~\ref{sec:saliency_models}, based on an informal assessment of their likely quality and on the availability of data or source code. We either obtained source code from the authors or implemented the method proposed in their papers:

\paragraph*{Saliency of large point sets} ($\mathbf{LS}$) \cite{Shtrom2013}:
We use our implementation of Shtrom et al.'s technique~\cite{Shtrom2013} to generate the saliency maps used in the evaluation. 

\paragraph*{Mesh saliency via spectral processing} ($\mathbf{MS}$) \cite{Song2014}:
Source code provided by the authors. Note that the Laplacian on a range scan with isolated points is not guaranteed to be semi positive definite, and thus may not have real eigenvectors. In these cases, we assign a default saliency value of 0 to all points. 

\paragraph*{Cluster-based point set saliency} ($\mathbf{CS}$) \cite{Tasse2015}: Source code provided by the authors.

\paragraph*{PCA-based saliency} ($\mathbf{PS}$) We implemented this saliency model as described in Section~\ref{sec:pca_based_saliency}.

Table~\ref{table:params_per_method} gives the parameter values used for the four saliency methods above, and introduced in Section~\ref{sec:saliency_models}. 

In addition to the above computational saliency models, we evaluate the following two baseline models:
\paragraph*{Chance} ($\mathbf{RS}$): We test saliency performance when saliency values are randomly assigned. Computational saliency models should have a better performance than a random model. For each point on a mesh, we choose at random a value between $0.0$ and $1.0$ to set its saliency value. 

\begin{figure*}[t]
\centering
\includegraphics[width=0.33\linewidth]{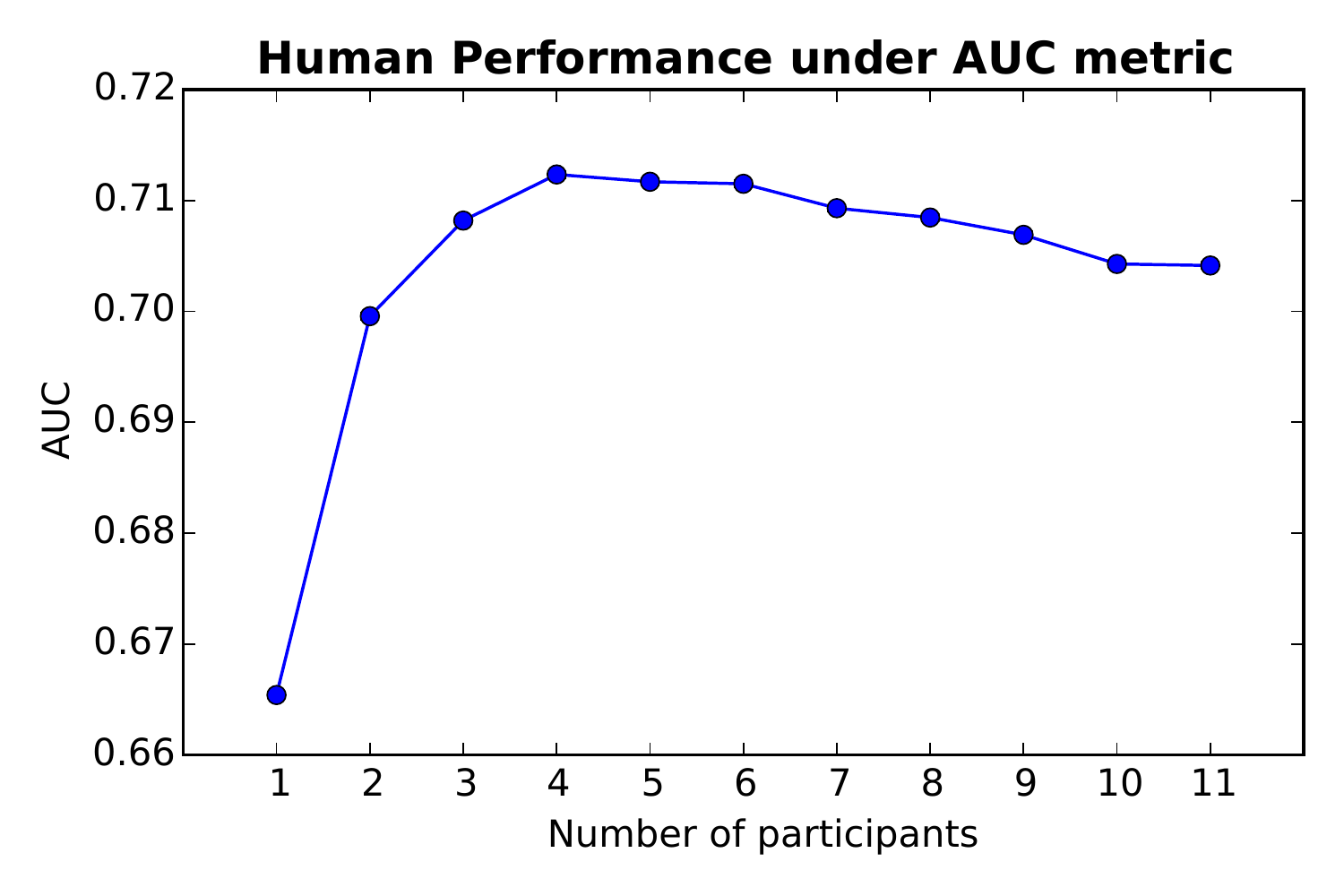}
\includegraphics[width=0.33\linewidth]{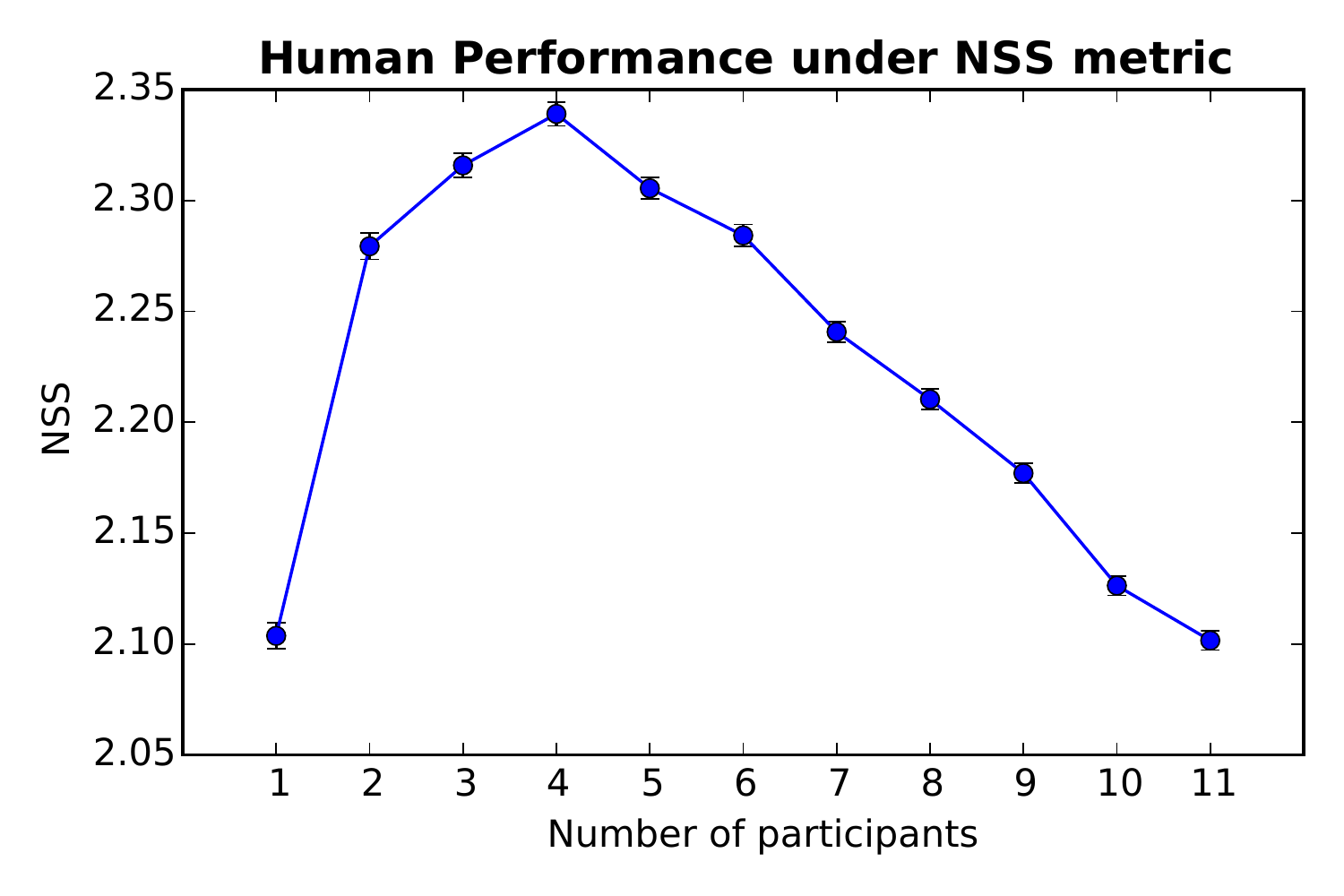}
\includegraphics[width=0.33\linewidth]{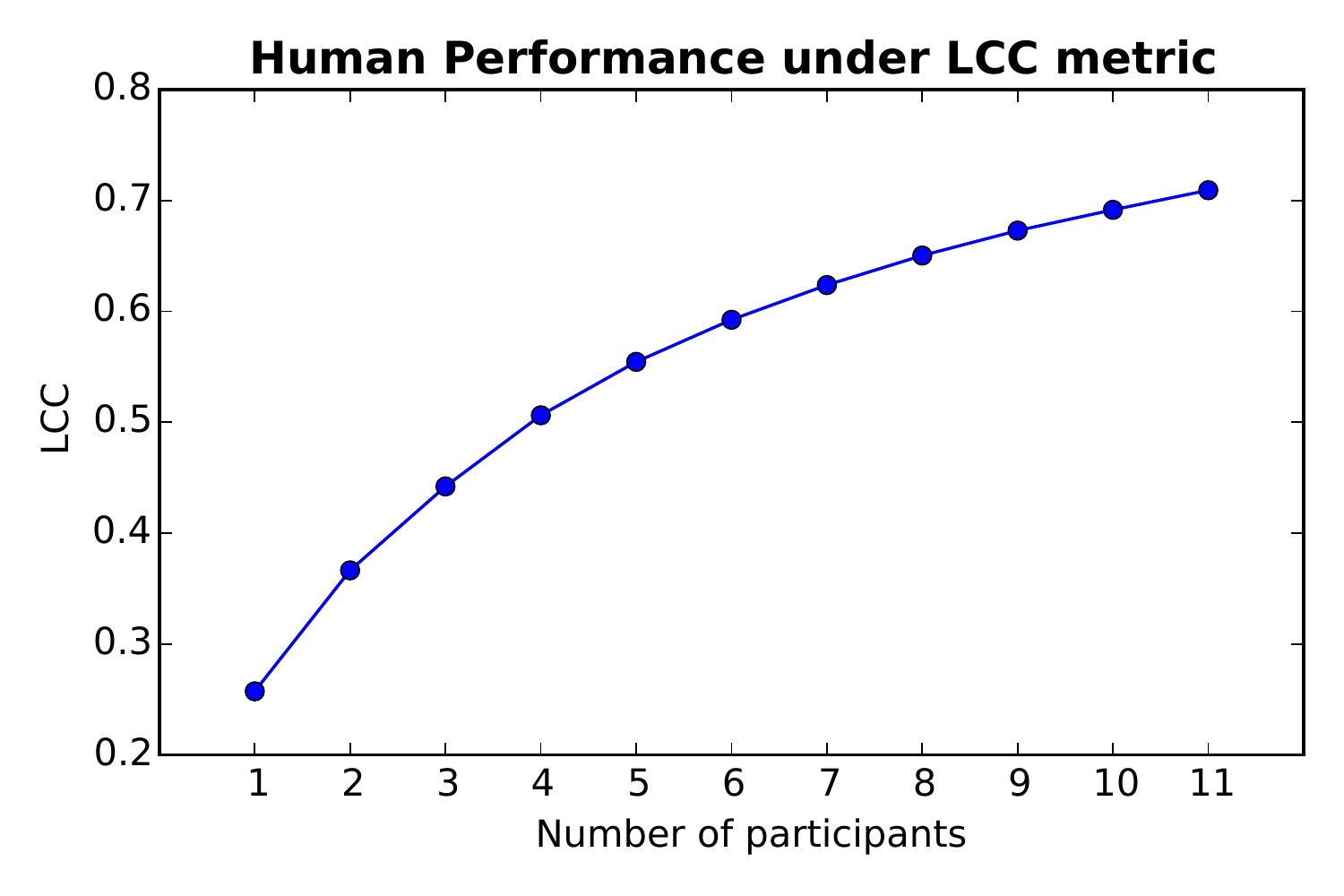}
\caption{Performance of $n_p$ participants to predict saliency from $n_p$ participants, as $n_p$ varies from $1$ to $11$.}

\label{fig:human_performance}
\end{figure*}

\paragraph*{Human performance} ($\mathbf{HS}$): Here, we are interested in how well one or more human participants' predictions differ from the consensus of all the participants. We investigate how well saliency data collected from $n_p$ participants predict ground-truth saliency from other participants. By default, we use $n_p=1$ but we vary $n_p$ to produce the results in Figure~\ref{fig:human_performance}. We can say that a computational saliency model predicts saliency as well as a human, if its performance is similar to the performance of the average human.

The six saliency models we are evaluating  vary significantly in how saliency is distributed over the mesh, with some models having more salient regions than others. This difference in the distributions of saliency maps is illustrated in Figure~\ref{fig:saliency_maps_avg_distribution}. The distribution of HS is equivalent to that of GS when summed up over  all participants. For a fair comparison, the histogram of each saliency map is matched to that of the ground-truth, similarly to the previous work on 2D saliency evaluation \cite{Judd2012}. We refer to the average histogram of a ground-truth saliency map as the reference histogram.  Given a saliency map, we find the mapping that optimally transforms its cumulative distribution function towards that of the reference histogram \cite[Section~5.6.4]{Gonzalez2001}. This ensures that all saliency maps have the same distribution of saliency values.  Examples of saliency maps with histograms matched with ground-truth are displayed in Figure~\ref{fig:maps_after_matching_hist}.

\begin{table}[t]
\centering
\begin{tabular}{lr@{\,=\,}lr@{\,=\,}l}
\multicolumn{1}{l}{Method} & \multicolumn{4}{c}{Parameters}\\
\hline
$\mathbf{LS}$ \cite{Shtrom2013} & $r_{low}$&$0.01R$ & $r_{high}$&$0.1R$ \\ % Saliency of large point sets (
$\mathbf{MS}$ \cite{Song2014} & $\epsilon$&$0.004R$ & $n$&$9$ \\ % Mesh saliency via spectral processing (
$\mathbf{CS}$ \cite{Tasse2015} & $r$&$0.02R$ & $K$&$100$ \\ % Cluster-based point set saliency (
$\mathbf{PS}$  & $r$&$0.01R$ \\ % PCA-based saliency  (
\hline
\end{tabular}
\caption{Parameters per evaluated saliency method. We used parameter values recommended in the corresponding papers and presented in Section~\ref{sec:saliency_models}. $R$ is the bounding sphere radius.}\label{table:params_per_method} %, with the exception of PS which was introduced in this paper.
\end{table}

\section{Experimental results}\label{sec:benchmark_results}

\begin{figure*}[!p]
\centering
\vspace{-2 mm}

\includegraphics[width=0.64\linewidth]{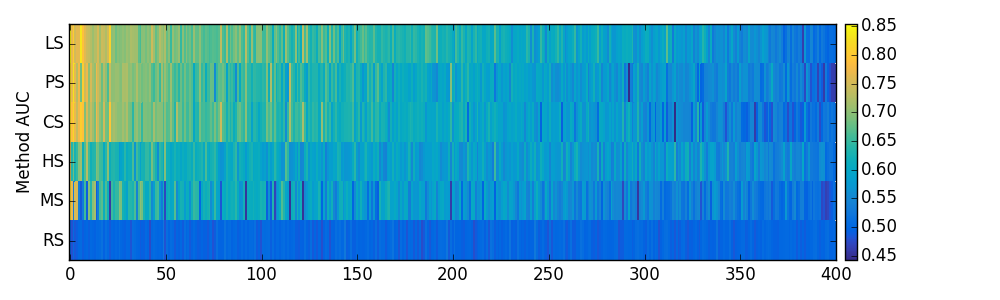}
\vspace{-2 mm}
\includegraphics[width=0.253\linewidth]{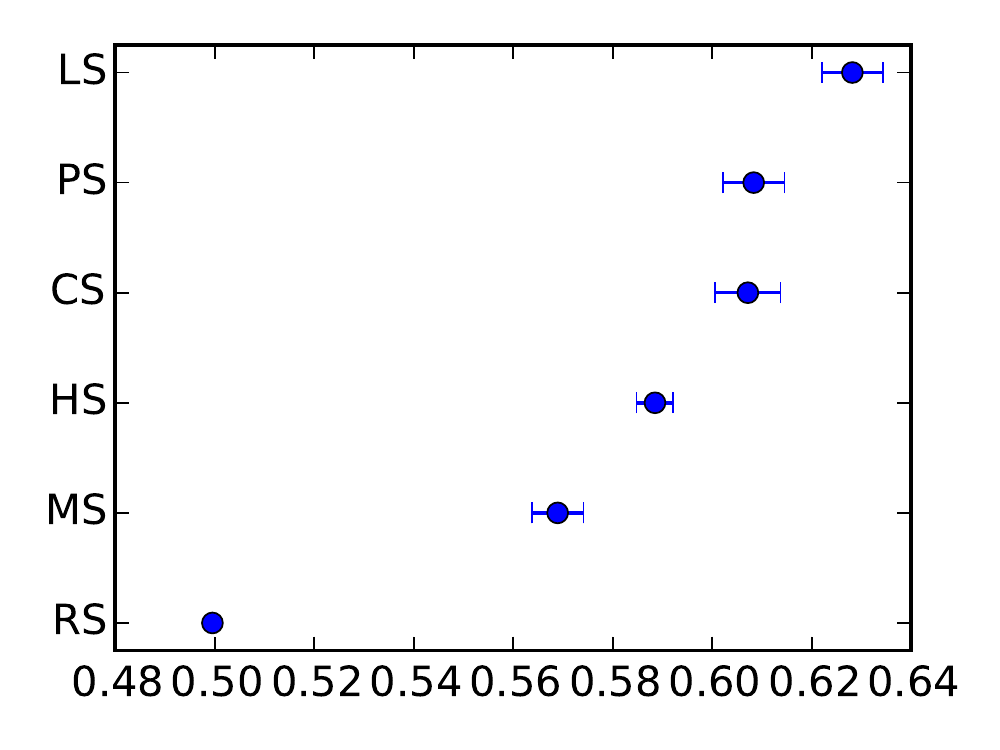}
\begin{minipage}[c]{0.69\linewidth}
\includegraphics[trim=-32 25.5 0 0, clip, width=0.835\linewidth]{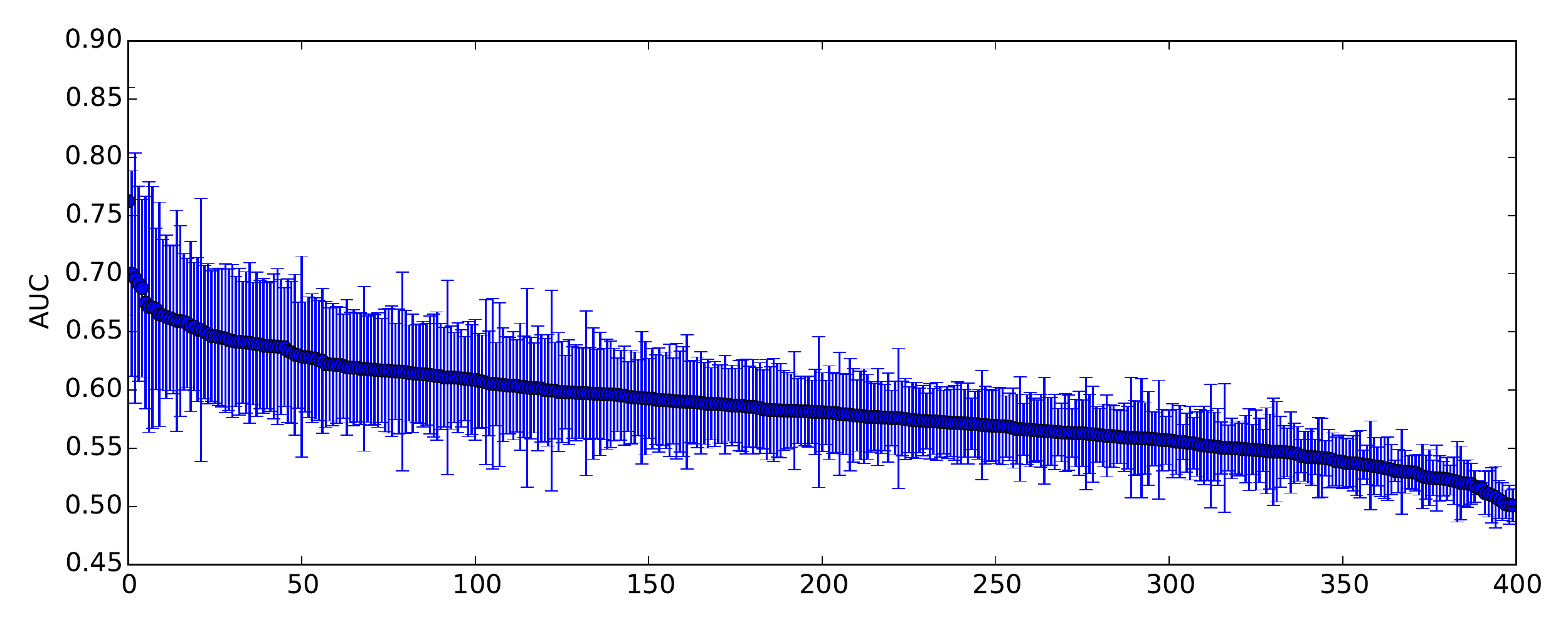}
\end{minipage}
\begin{minipage}[c]{0.253\linewidth}
\includegraphics[width=0.44\linewidth]{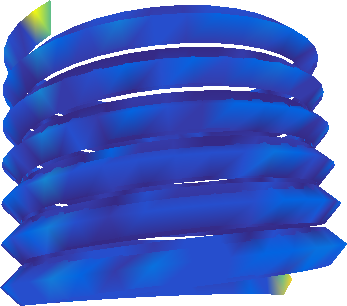}
\includegraphics[width=0.44\linewidth]{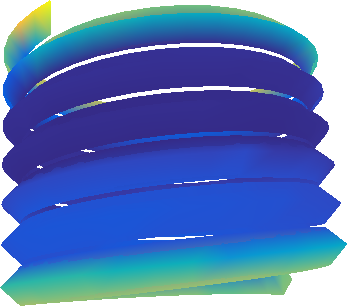}\hfill
\includegraphics[width=0.07\linewidth]{eval_results/parula_colorbar.png}
\end{minipage}
\rule[1ex]{\linewidth}{0.5pt}

\vspace{-1 mm}
\includegraphics[width=0.64\linewidth]{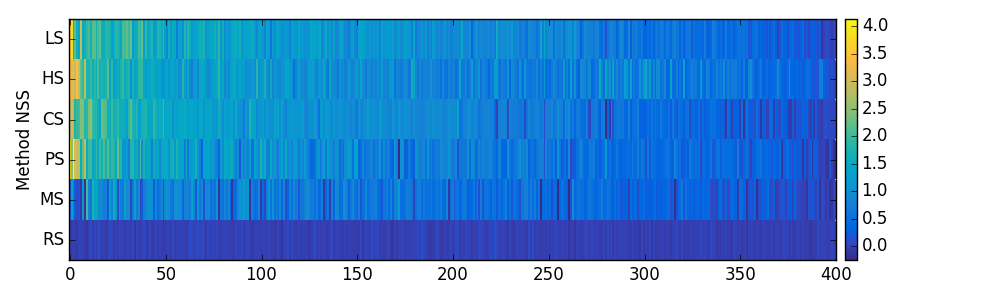}
\vspace{-2.5 mm}
\includegraphics[width=0.253\linewidth]{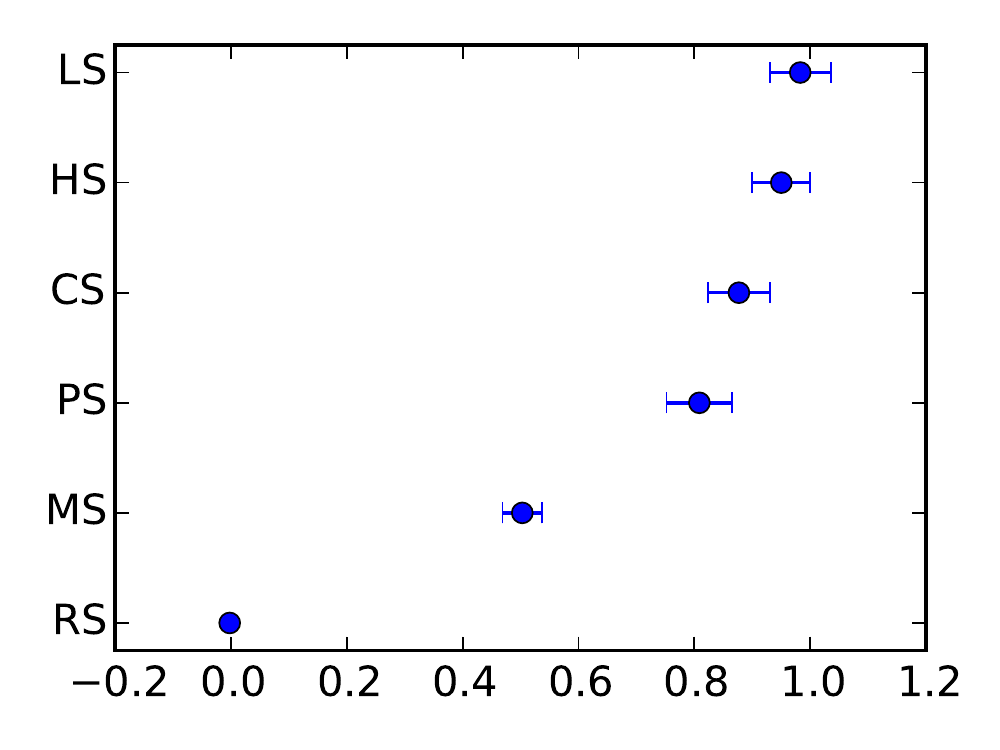}
\begin{minipage}{0.71\linewidth}
\includegraphics[trim=-44 25.5 0 0, clip, width=0.825\linewidth]{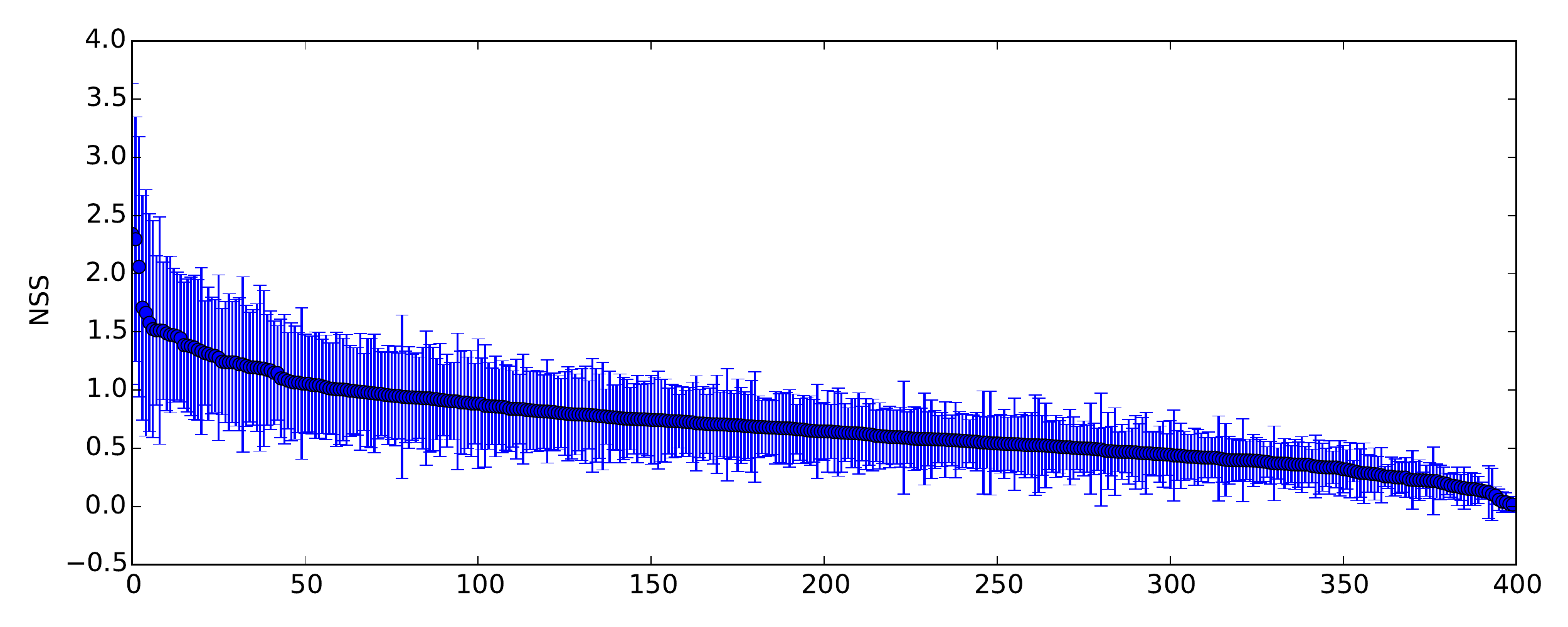}
\end{minipage}
\begin{minipage}[c]{0.253\linewidth}
\centering
\includegraphics[width=0.44\linewidth]{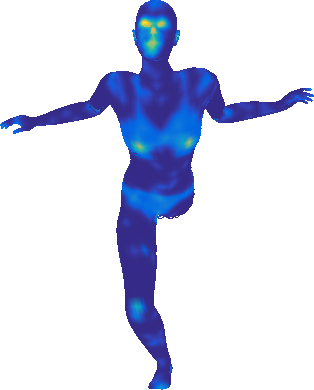}
\includegraphics[width=0.44\linewidth]{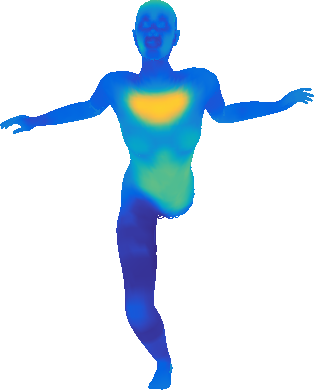}\hfill
\includegraphics[width=0.07\linewidth]{eval_results/parula_colorbar.png}
\end{minipage}
\rule[1ex]{\linewidth}{0.5pt}

\vspace{-1 mm}
\includegraphics[width=0.64\linewidth]{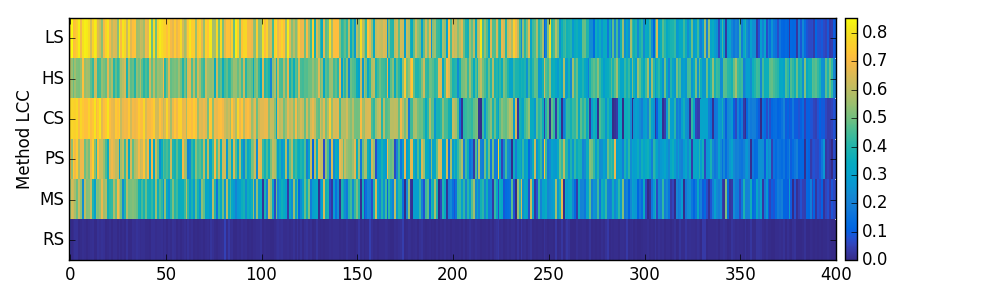}
\vspace{-2.5 mm}
\includegraphics[width=0.253\linewidth]{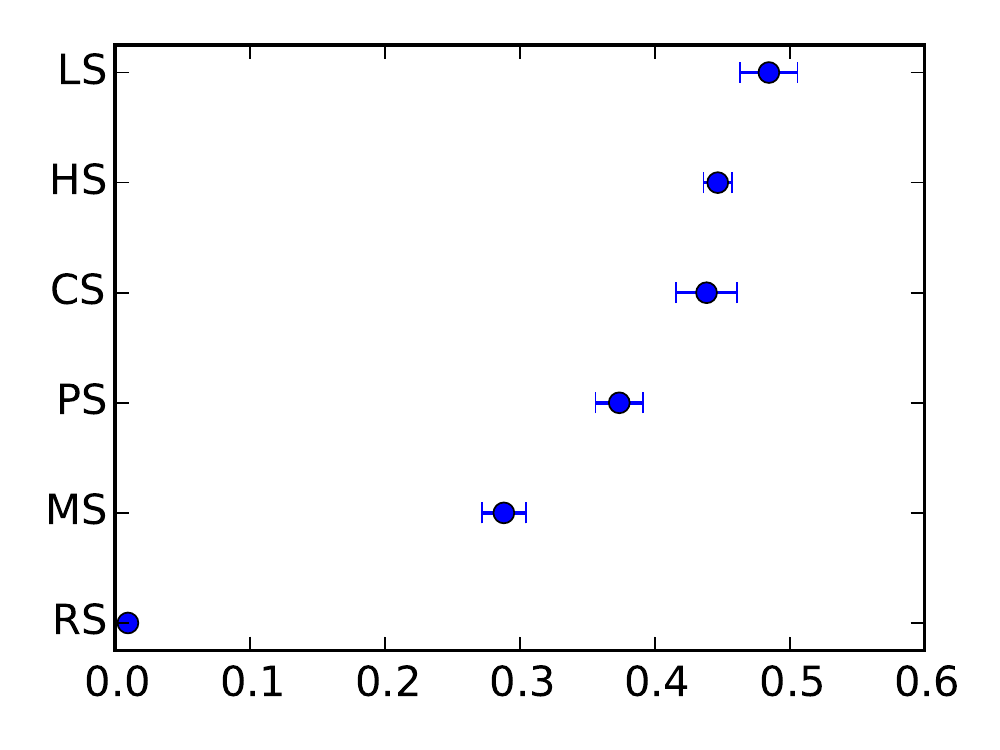}
\begin{minipage}{0.71\linewidth}
\includegraphics[trim=-43 25 0 0, clip, width=0.825\linewidth]{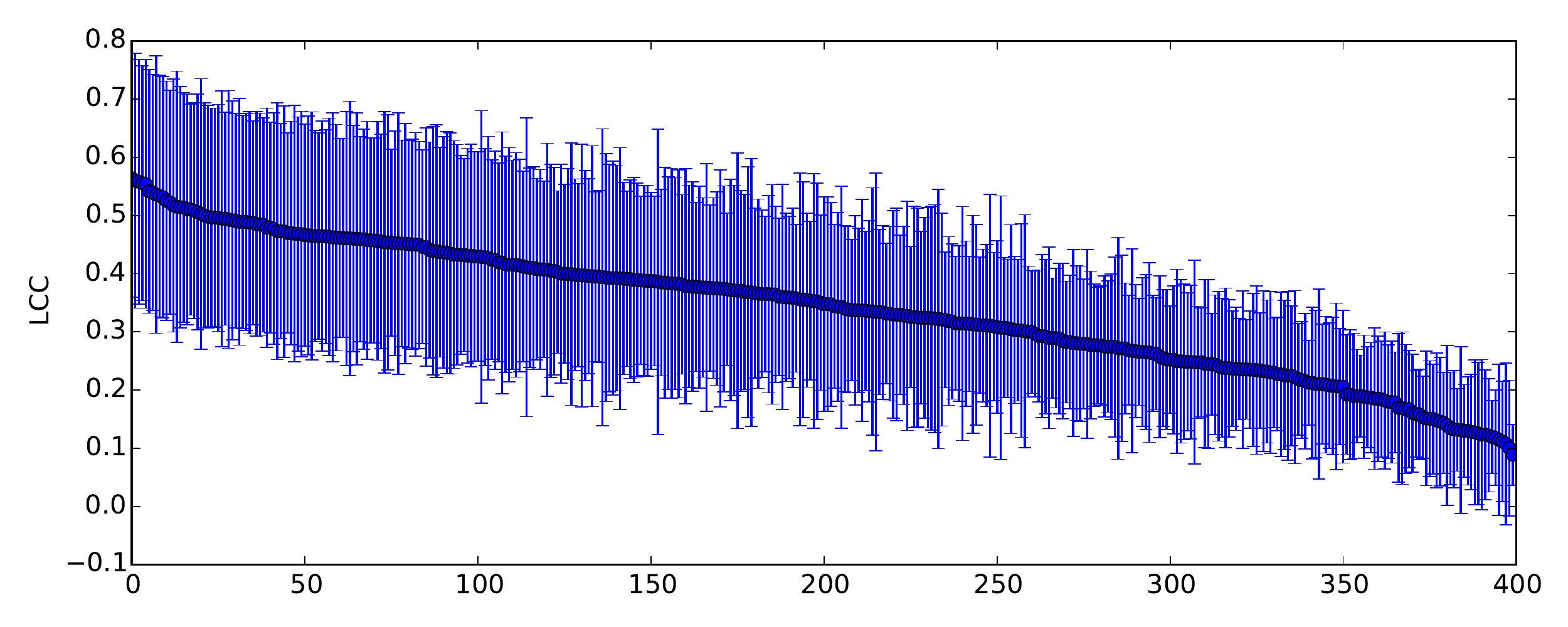}
\end{minipage}
\begin{minipage}[c]{0.253\linewidth}
\centering
\includegraphics[width=0.34\linewidth]{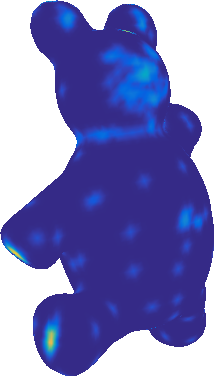}
\includegraphics[width=0.34\linewidth]{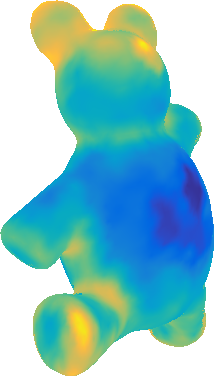}\hfill
\includegraphics[width=0.07\linewidth]{eval_results/parula_colorbar.png}
\end{minipage}

\caption{Saliency performance evaluation. The figure has three parts, one for each metric: AUC (top), NSS (middle), and LCC (bottom). Each part contains four panels. Top-left: a color map with the x-axis showing shapes ordered by their average score and the color scale representing the actual metric value. Top-right: saliency models ordered by their average score, with blue error bars representing 95\% confidence intervals. Bottom-left: shape mean saliency performance with shapes ordered by their average score. Bottom-right: ground-truth and top-performing saliency maps for shapes with poor average scores, illustrating saliency on `difficult' shapes. The panel associated with the NSS metric shows the shape with the second lowest score because both AUC and NSS have the same worst-performing shape.}
\label{fig:saliency_performance}
\end{figure*}

\begin{table*}[!ht]
\centering
\setlength{\tabcolsep}{1.5pt}
\caption{AUC performance per shape class in SHREC07. For each class, we average the saliency AUC scores of all shapes belonging to that class. The last column
reports the average of all seven AUC scores. The table shows, for instance, that it is easier to detect saliency on a \emph{mechanic} shape than a \emph{spring}. It also shows that no tested saliency model can yet detect saliency on a \emph{teddy} shape better than a human.}
\label{auc_per-class}
\begin{tabular}{lcccccccc}
\hline
Class      & LS (Shtrom2013)          & PS (PCA-based)           & CS (Tasse2015)  & HS (1 vs all)            & MS (Song2014)   & RS (Random)     & Avg-AUC         \\
\hline
\emph{mechanic} & $0.69 \pm 0.03$ & $\mathbf{0.71 \pm 0.03}$ & $0.70 \pm 0.03$ & $0.64 \pm 0.02$ & $0.53 \pm 0.03$ & $0.50 \pm 0.00$ & $0.63 \pm 0.05$ \\
\emph{fish} & $0.67 \pm 0.02$ & $\mathbf{0.67 \pm 0.02}$ & $0.67 \pm 0.01$ & $0.61 \pm 0.01$ & $0.63 \pm 0.01$ & $0.50 \pm 0.00$ & $0.62 \pm 0.03$ \\
\emph{armadillo} & $\mathbf{0.67 \pm 0.01}$ & $0.66 \pm 0.01$ & $0.66 \pm 0.01$ & $0.62 \pm 0.01$ & $0.60 \pm 0.01$ & $0.50 \pm 0.00$ & $0.62 \pm 0.03$ \\
\emph{airplane} & $\mathbf{0.67 \pm 0.02}$ & $0.64 \pm 0.02$ & $0.63 \pm 0.02$ & $0.60 \pm 0.01$ & $0.62 \pm 0.02$ & $0.50 \pm 0.00$ & $0.61 \pm 0.03$ \\
\emph{table} & $\mathbf{0.67 \pm 0.03}$ & $0.62 \pm 0.02$ & $0.63 \pm 0.03$ & $0.61 \pm 0.02$ & $0.58 \pm 0.02$ & $0.50 \pm 0.00$ & $0.60 \pm 0.03$ \\
\emph{vase} & $0.62 \pm 0.02$ & $\mathbf{0.63 \pm 0.02}$ & $0.62 \pm 0.02$ & $0.58 \pm 0.01$ & $0.61 \pm 0.02$ & $0.50 \pm 0.00$ & $0.59 \pm 0.02$ \\
\emph{buste} & $0.63 \pm 0.03$ & $\mathbf{0.64 \pm 0.03}$ & $0.62 \pm 0.03$ & $0.59 \pm 0.01$ & $0.56 \pm 0.02$ & $0.50 \pm 0.00$ & $0.59 \pm 0.03$ \\
\emph{four-legged} & $\mathbf{0.62 \pm 0.03}$ & $0.61 \pm 0.03$ & $0.60 \pm 0.03$ & $0.59 \pm 0.01$ & $0.60 \pm 0.03$ & $0.50 \pm 0.00$ & $0.59 \pm 0.03$ \\
\emph{cup} & $\mathbf{0.62 \pm 0.02}$ & $0.61 \pm 0.02$ & $0.61 \pm 0.02$ & $0.57 \pm 0.01$ & $0.59 \pm 0.03$ & $0.50 \pm 0.00$ & $0.58 \pm 0.03$ \\
\emph{hand} & $\mathbf{0.63 \pm 0.02}$ & $0.60 \pm 0.02$ & $0.61 \pm 0.02$ & $0.58 \pm 0.01$ & $0.58 \pm 0.01$ & $0.50 \pm 0.00$ & $0.58 \pm 0.02$ \\
\emph{chair} & $\mathbf{0.66 \pm 0.02}$ & $0.58 \pm 0.03$ & $0.59 \pm 0.03$ & $0.62 \pm 0.02$ & $0.55 \pm 0.02$ & $0.50 \pm 0.00$ & $0.58 \pm 0.03$ \\
\emph{ant} & $\mathbf{0.63 \pm 0.01}$ & $0.58 \pm 0.01$ & $0.61 \pm 0.01$ & $0.60 \pm 0.01$ & $0.57 \pm 0.01$ & $0.50 \pm 0.00$ & $0.58 \pm 0.02$ \\
\emph{bearing} & $0.64 \pm 0.03$ & $0.63 \pm 0.03$ & $\mathbf{0.65 \pm 0.03}$ & $0.55 \pm 0.01$ & $0.50 \pm 0.02$ & $0.50 \pm 0.00$ & $0.58 \pm 0.04$ \\
\emph{bird} & $\mathbf{0.62 \pm 0.02}$ & $0.60 \pm 0.02$ & $0.60 \pm 0.02$ & $0.57 \pm 0.01$ & $0.56 \pm 0.02$ & $0.50 \pm 0.00$ & $0.58 \pm 0.02$ \\
\emph{plier} & $\mathbf{0.62 \pm 0.01}$ & $0.58 \pm 0.01$ & $0.60 \pm 0.01$ & $0.56 \pm 0.01$ & $0.56 \pm 0.01$ & $0.50 \pm 0.00$ & $0.57 \pm 0.02$ \\
\emph{human} & $0.59 \pm 0.02$ & $\mathbf{0.59 \pm 0.02}$ & $0.57 \pm 0.02$ & $0.58 \pm 0.01$ & $0.57 \pm 0.02$ & $0.50 \pm 0.00$ & $0.57 \pm 0.02$ \\
\emph{octopus} & $\mathbf{0.62 \pm 0.02}$ & $0.57 \pm 0.02$ & $0.55 \pm 0.02$ & $0.60 \pm 0.01$ & $0.53 \pm 0.01$ & $0.50 \pm 0.00$ & $0.56 \pm 0.02$ \\
\emph{teddy} & $0.56 \pm 0.01$ & $0.57 \pm 0.01$ & $0.57 \pm 0.01$ & $\mathbf{0.60 \pm 0.01}$ & $0.55 \pm 0.01$ & $0.50 \pm 0.00$ & $0.56 \pm 0.02$ \\
\emph{glasses} & $\mathbf{0.57 \pm 0.01}$ & $0.55 \pm 0.01$ & $0.52 \pm 0.02$ & $0.56 \pm 0.01$ & $0.53 \pm 0.01$ & $0.50 \pm 0.00$ & $0.54 \pm 0.02$ \\
\emph{spring} & $\mathbf{0.55 \pm 0.01}$ & $0.53 \pm 0.02$ & $0.55 \pm 0.02$ & $0.54 \pm 0.01$ & $0.55 \pm 0.01$ & $0.50 \pm 0.00$ & $0.54 \pm 0.02$ \\
%\hline
\end{tabular}
\end{table*}

We present the performance of the six selected saliency models, under the AUC, NSS and LCC metrics. We also present results on how human performance varies with the number of participants, and human consistency.

\subsection{Saliency on watertight meshes}
First, we report results on the SHREC07 dataset.

\subsubsection{Model performances}\label{sec:shre07_model_performances}

We discuss the performance results illustrated in Figure~\ref{fig:saliency_performance}. 

\paragraph*{AUC}is the area under the ROC curve generated by plotting true positive rate against false positive rate. Figure~\ref{fig:saliency_performance} (top) shows the performance of the selected saliency models, under the AUC metric. All models are significantly better, on average, than chance. Saliency models based on the FPFH descriptor all perform better than both human performance HS and spectral mesh saliency MS. Among these descriptor-based methods, LS has the highest AUC performance. There is no statistically significant difference between the other two descriptor-based techniques PS and  CS.

\paragraph*{NSS}is the average of saliency values at human-selected keypoints.  Figure~\ref{fig:saliency_performance} (middle) presents models' performance under this metric. Similarly to the AUC metric, all models perform better than chance under NSS, with spectral mesh saliency MS having a statistically lower mean score than others. LS performs as well as human performance, with no statistically significant difference between their two mean scores. CS is not statistically different from PS. Note that under the NSS metric, HS is one of the top performing models, in contrast with its low AUC score. This is because AUC is influenced by the ordering of saliency values within a ground-truth saliency map. NSS uses the user-selected keypoints directly in its formulation with no consideration of how many times a point was selected by participants.

\paragraph*{LCC}estimates the strength of the relationship between the distributions of a shape saliency map and its ground-truth saliency. Results of models' performance under LCC are presented in Figure~\ref{fig:saliency_performance} (bottom). Models' rankings under this metric are similar to rankings by NSS scores. The key difference is that there is no significant difference between CS and human performance HS. Thus CS performs better under LCC.

In summary, LS \cite{Shtrom2013} achieves the best performance under all metrics. There is little difference between the performance of the other two FPFH-based saliency models, namely CS \cite{Tasse2015} and PCA-based saliency. It is important to note that the PCA approach to 3D saliency, first introduced in this paper, uses dramatically less data in its comparisons and yet produces competitive results compared to the state-of-the-art. 
Of course, while PS uses only a single value in its comparison (the first principal axis of the PCA result), it comes at the expense of a preprocessing step: running PCA on an $n\times33$ matrix to get the single saliency value, where $n$ is the number of points. This is reasonably fast and the preprocessing is amortized over many thousands of queries.

\subsubsection{Performance per shape classes}

We analyse how the selected saliency models perform for each of the $20$ classes in SHREC07. This is  interesting because it gives us  insight into which classes have the worst saliency detection and thus could benefit from future work in the field. Table~\ref{auc_per-class} shows AUC scores for each class and saliency model. For the class \emph{teddy}, human performance (HS) outperforms all other saliency models. In this particular case, human participants know that real-life shapes from this class have facial features and thus consider the face to be salient even if the 3D shape presented to them has no discriminating features. This is an example of humans using their prior experience in saliency detection, which is not yet possible for unsupervised  saliency models. Man-made shapes such as \emph{mechanic}, \emph{airplane} and \emph{chair} are easier for saliency detection due to their sharp features and simple structure.

\subsubsection{Human performance and consistency}
Figure~\ref{fig:human_performance} illustrates how well $n_p$ participants predict the ground-truth averaged over $n_p$ participants, under AUC, NSS and LCC. AUC performance stops improving significantly when $n_p$ reaches $3$ participants. NSS only increases significantly when $n_p$ increases to $2$ and starts to drop off for $n_p>4$. The drop is not statistically significant. 
In contrast with these metrics $LCC$ keeps improving with increasing number of participants. $LCC$ computes the correlation between the distributions of a saliency map and ground-truth saliency, which explains why this value increases with the number of participants. The negligible difference in AUC and NSS performance with $n_p \geq 3$ 
may suggest that on average $3$ participants are enough to obtain ground-truth saliency. 

\begin{figure*}[t]
\centering
\begin{tabular}{ccc}
\textbf{AUC}  & \textbf{NSS} & \textbf{LCC} \\
\includegraphics[width=0.32\textwidth]{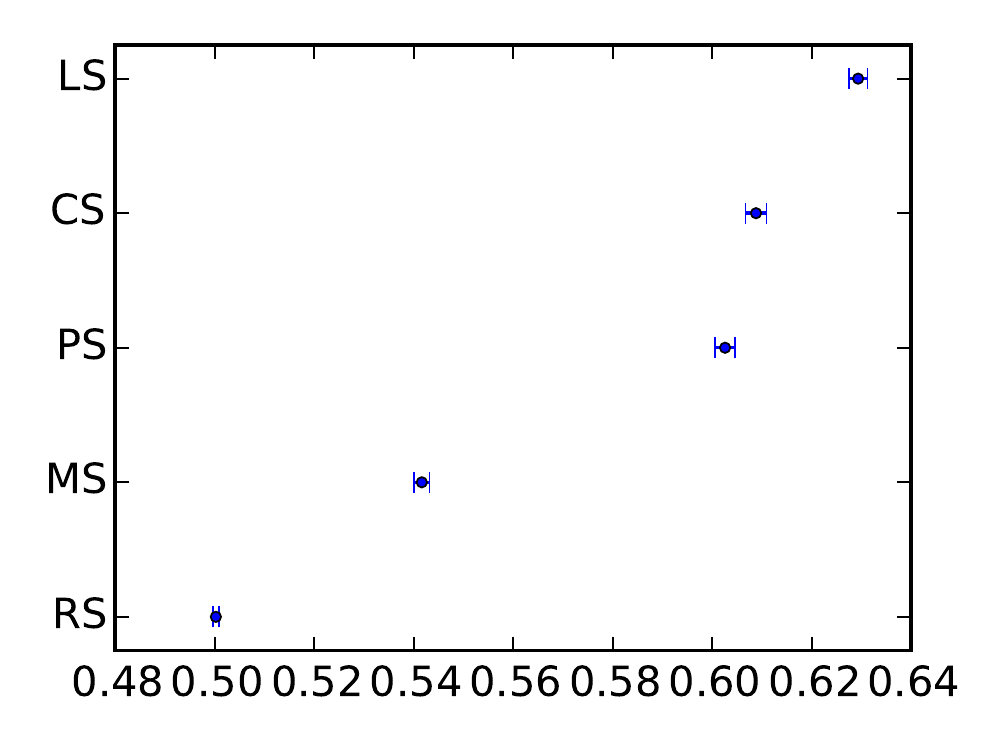} &
\includegraphics[width=0.32\textwidth]{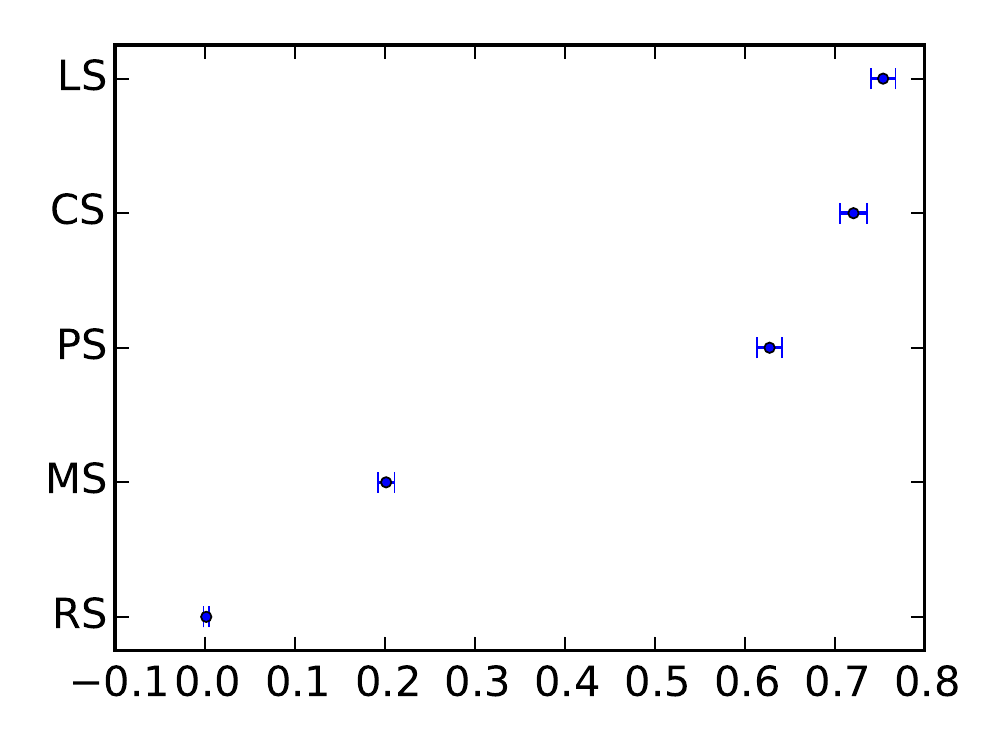} &
\includegraphics[width=0.32\textwidth]{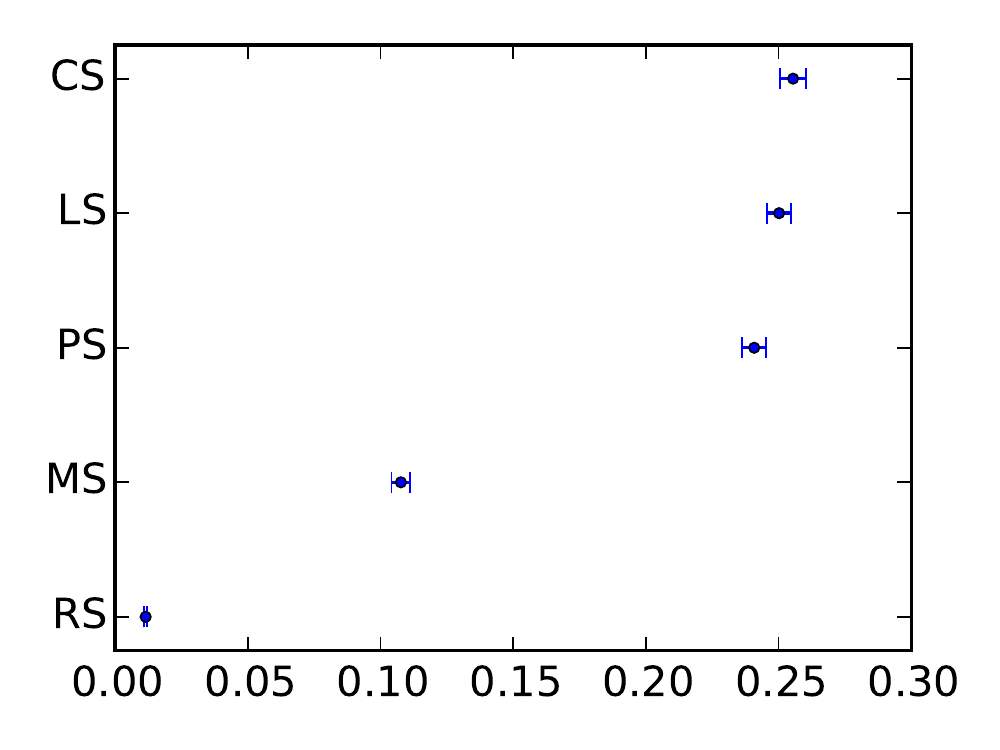} 
\end{tabular}
\caption{Saliency performance on simulated range scans, ordered by the average metric score per method. }\label{fig:scans_saliency_performance}
\end{figure*}

\subsection{Saliency on simulated range scans}\label{sec:scans_model_performances}
\begin{figure*}[p]
\centering
{
\begin{tabular}{ccccc}
Ground-truth \cite{Chen2012} & LS \cite{Shtrom2013} & MS \cite{Song2014} & CS \cite{Tasse2015} & PS (PCA-based) \\
\hline
\\
\includegraphics[width=\teaserfigwidth,angle=-90,origin=c]{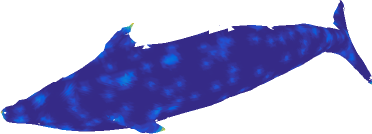} &
\includegraphics[width=\teaserfigwidth,angle=-90,origin=c]{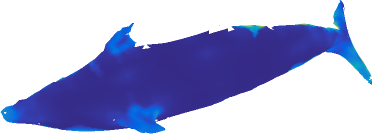} &
\includegraphics[width=\teaserfigwidth,angle=-90,origin=c]{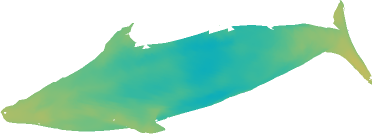} &
\includegraphics[width=\teaserfigwidth,angle=-90,origin=c]{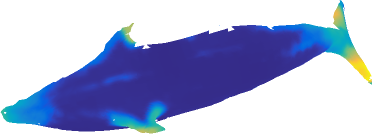} &
\includegraphics[width=\teaserfigwidth,angle=-90,origin=c]{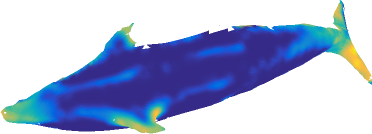} \\
(AUC, NSS, LCC): & (0.61, \textbf{0.44}, 0.13) & (0.57, 0.21, 0.05) & (\textbf{0.62}, 0.41, 0.18) & (0.61, 0.40, \textbf{0.18}) \\

\includegraphics[width=\teaserfigwidth]{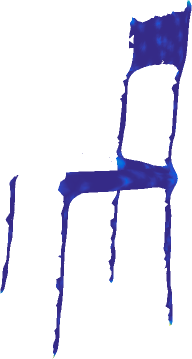} &
\includegraphics[width=\teaserfigwidth]{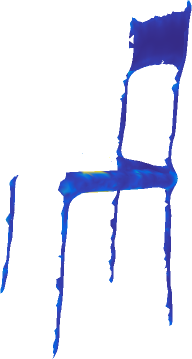}  &
\includegraphics[width=\teaserfigwidth]{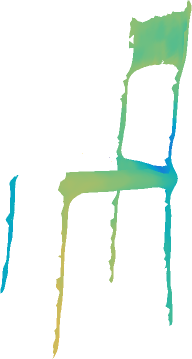} &
\includegraphics[width=\teaserfigwidth]{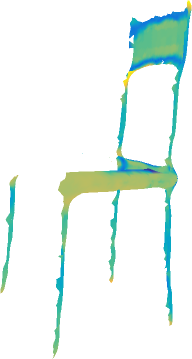} &
\includegraphics[width=\teaserfigwidth]{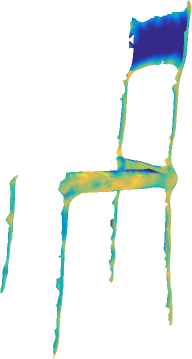} \\
(AUC, NSS, LCC): & (\textbf{0.65}, 0.45, 0.18) & (0.62, 0.67, \textbf{0.39}) & (0.63, 0.44, 0.09) & (0.64, \textbf{0.89}, 0.28) \\

\includegraphics[width=\teaserfigwidth]{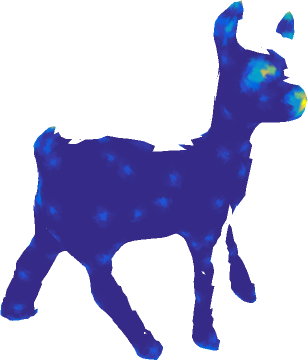}  &
\includegraphics[width=\teaserfigwidth]{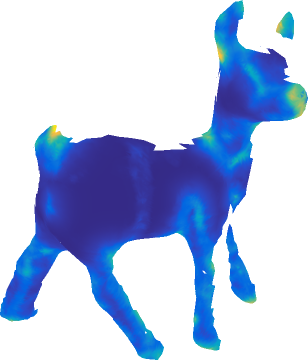}  &
\includegraphics[width=\teaserfigwidth]{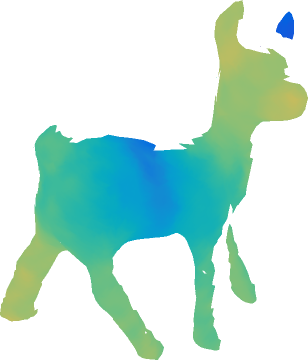} &
\includegraphics[width=\teaserfigwidth]{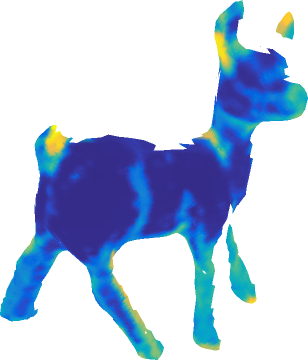} &
\includegraphics[width=\teaserfigwidth]{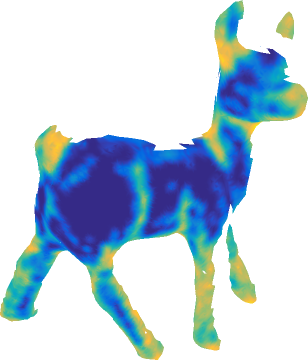} \\
(AUC, NSS, LCC): & (\textbf{0.69}, \textbf{1.21}, \textbf{0.35}) & (0.60, 0.41, 0.25) & (0.62, 0.81, 0.18) & (0.65, 0.63, 0.11) \\

\includegraphics[width=\teaserfigwidth]{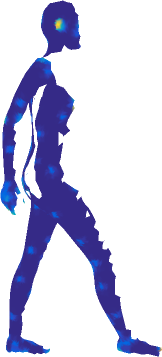} &
\includegraphics[width=\teaserfigwidth]{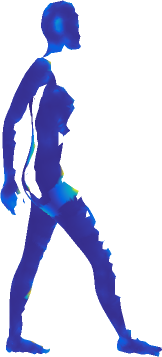} &
\includegraphics[width=\teaserfigwidth]{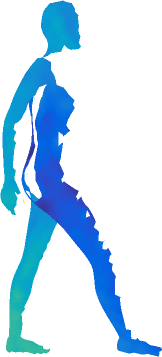} &
\includegraphics[width=\teaserfigwidth]{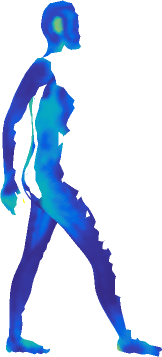} &
\includegraphics[width=\teaserfigwidth]{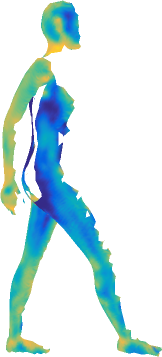} \\
(AUC, NSS, LCC): & (\textbf{0.61}, 0.75, 0.18) & (0.53, 0.01, 0.08) & (0.61, \textbf{0.77}, \textbf{0.33}) & (0.57, 0.31, 0.10) \\
\end{tabular}
}
\caption{Saliency maps (before histogram matching) on 4 simulated range scans, obtained by virtually scanning watertight meshes in the SHREC07 dataset.}\label{fig:scans_saliency}
\end{figure*}

We present saliency performance on range scans generated from the SHREC07 dataset used above. Figure~\ref{fig:scans_saliency_performance} presents average performances per metric and Figure~\ref{fig:scans_saliency} illustrates saliency maps on a few scans. We did not include HS in the evaluated methods for this simulated dataset since the human participants did not directly work with range scans.

\paragraph*{AUC} Similarly to results on SHREC07, LS has the top AUC performance. All other performance differences are significant: CS is better than PS, which performs better than MS. Thus, CS performs similarly to PS on the set of watertight meshes, but outperforms the latter on range scans.

\paragraph*{NSS}  LS is the top performing method, followed by CS and PS. Again, the difference between CS and PS is only significant on range scans. MS and RS have the lowest NSS scores. This is because MS is  penalized for its spectral-based approach that is designed for 2D manifolds and is not able to support any range scan.

\paragraph*{CC} All performance differences between methods were significant, with the exception of CS and LS. This reinforces the indication that CS compares better with other methods when computed over range scans.

In summary, LS remains the top performing method on 2D manifolds and range scans. It is followed by CS, another method designed for point sets. MS performs poorly (although better than RS) on range scans since it requires a 2D manifold to obtain real solutions of the Laplacian.

\section{Discussion}

This paper reported an analysis of saliency models on watertight meshes and range scans. We did not report timings since implementations were built in different programming environments, including C++, Matlab and Python. 

Results show, in Section~\ref{sec:shre07_model_performances}, that the two-scale approach of LS appears to detect saliency peaks similar to ground-truth irrespective of the data type. The other multi-scale approach MS has the lowest performance, after chance, possibly due to the fact that it was designed to capture both large and small salient regions, while the ground-truth collection process is focused on local salient patches. An advantage of MS  is that it is invariant to isometric transformations, since the Laplacian is an intrinsic operator. We expect that on a dataset of non-rigid shapes MS will have better performance compared to the alternatives. This would be an interesting investigation for future work.

We also introduce a simple saliency method based on PCA analysis of FPFH descriptors. Results on watertight meshes show that it outperforms HS, and performs similarly to CS on some metrics. On range scans, PS only performed better than MS and RS although the differences with LS and CS were not large. Given the simplicity of the PCA-based saliency and the small performance loss compared to better saliency models, it may be a good candidate for applications where accuracy loss in the saliency map can be tolerated. Examples of such applications are shape simplifications \cite{Song2014} and viewpoint selection \cite{Shtrom2013}.

Our evaluation against ground-truth is based on saliency maps that were extracted from multiple human-selected keypoints. Such saliency maps do not accurately reflect the result of low-level visual attention. Running a similar quantitative analysis on eye fixations data will help determine which methods are closer to emulating low-level saliency.  

\section{Conclusion}
We introduce the first saliency evaluation framework for 3D shapes, based on three performance metrics. We compare six saliency models including  chance, human performance and a PCA-based approach. The performance results showed that all models performed better than chance, and Shtrom et al.'s method \cite{Shtrom2013} has, on average, the best scores on both watertight meshes and simulated range scans. PCA-based saliency performed as well as  cluster-based point set saliency \cite{Tasse2015}, and significantly better than spectral mesh saliency~\cite{Song2014}.  We are releasing our benchmark online, so that previous methods not tested here and future saliency techniques can be evaluated objectively. 
%This fills a gap in the saliency literature and will help the community judge the success of new methods.

\bibliographystyle{eg-alpha-doi}
\bibliography{paper}

\end{document}